\RequirePackage{fixltx2e,textcomp,graphicx,amssymb,amsmath}

\documentclass[10pt]{iopart} 

\usepackage[english]{babel}

\begin{document}
\title[Architecture of nanoscale ferroelectric domains in GaMo$_4$S$_8$]{Architecture of nanoscale ferroelectric domains in GaMo$_4$S$_8$}
\author{
Erik~Neuber\textsuperscript{1},
Peter~Milde\textsuperscript{1} \footnote{Corresponding author: Peter.Milde@tu-dresden.de},
Adam~Butykai\textsuperscript{2},
Sandor~Bordacs\textsuperscript{2},
Hiroyuki~Nakamura\textsuperscript{3},
Takeshi~Waki\textsuperscript{3},
Yoshikazu~Tabata\textsuperscript{3},
Korbinian~Geirhos\textsuperscript{4},
Peter~Lunkenheimer\textsuperscript{4},
Istvan~K\'{e}zsm\'{a}rki\textsuperscript{2,4},
Petr~Ondrejkovic\textsuperscript{5},
Jirka~Hlinka\textsuperscript{5},
and Lukas~M.~Eng\textsuperscript{1,6}
}

\address{\textsuperscript{1} Institute of Applied Physics, Technische Universit\"{a}t Dresden, D-01062 Dresden, Germany}
\address{\textsuperscript{2} Department of Physics, Budapest University of Technology and Economics and MTA-BME Lend\"{u}let Magneto-optical Spectroscopy Research Group, 1111 Budapest, Hungary}
\address{\textsuperscript{3} Department of Materials Science and Engineering, Kyoto University, Kyoto 606-8501, Japan}
\address{\textsuperscript{4} Experimental Physics V, Center for Electronic Correlations and Magnetism, University of Augsburg, 86135 Augsburg, Germany}
\address{\textsuperscript{5} Institute of Physics, Academy of Sciences of the Czech Republic, Na Slovance 2, 18221, Prague, Czech Republic}
\address{\textsuperscript{6} Center for Advancing Electronics Dresden (cfaed), Technische Universit\"at Dresden, D-01062 Dresden, Germany}

\ead{Peter.Milde@tu-dresden.de}
\begin{abstract}
Local-probe imaging of the ferroelectric domain structure and auxiliary bulk pyroelectric measurements were conducted at low temperatures with the aim to clarify the essential aspects of the orbitally driven phase transition in GaMo\textsubscript{4}S\textsubscript{8}, a lacunar spinel crystal that can be viewed as a spin-hole analogue of its GaV\textsubscript{4}S\textsubscript{8} counterpart.
We employed multiple scanning probe techniques combined with symmetry and mechanical compatibility analysis to uncover the hierarchical domain structures, developing on the 10-100~nm scale.
The identified domain architecture involves a plethora of ferroelectric domain boundaries and junctions, including primary and secondary domain walls in both electrically neutral and charged configurations, and topological line defects transforming neutral secondary walls into two oppositely charged ones.
\end{abstract}
\noindent \textit{Kelvin probe force microscopy, piezoresponse force microscopy, lacunar spinel, ferroelectricity, multiferroicity, domain walls, topology}
\maketitle
\ioptwocol 

\section{Introduction}
Interplay between ferroelectricity and magnetism in multiferroics holds an enormous potential for new-generation magnetic memories and spintronic devices \cite{Fiebig:JoPD-2005-38-8,Spaldin:Sci-2005-309-5733}, as it enables the manipulation of magnetic states via electric fields without Joule heating.
The recent observation of magnetic skyrmions -- non-collinear spin patterns with a non-trivial topology and typical size ranging from a few nanometers up to a few hundreds of nanometers -- also opens new perspectives for these applications \cite{Muhlbauer:Sci-2009-323-5916, Fert:NatNano-2013-8-3, Romming:Sci-2013-341-6164, Kezsmarki:NatM-2015-14, Yu:Nat-2010-465-7300, Milde:Sci-2013-340-6136, Seki:Sci-2012-336-6078, Bordacs:SR-2017-7-7584}.
In this respect, compounds being multiferroic and hosting magnetic skyrmions at the same time represent the ideal material class.

Lacunar spinels, including GaV\textsubscript{4}S\textsubscript{8} and GaV\textsubscript{4}Se\textsubscript{8}, have indeed been identified as multiferroic skyrmion host materials, showing electric polarization and exhibiting a N{\'e}el-type skyrmion lattice (SkL) state \cite{Kezsmarki:NatM-2015-14, Ruff:SciAdv-2015-1-10, Wang:PRL-2015-115-20, Ehlers:PRevB-2016-94-1, Hlinka:PRevB-2016-94-6}.
GaV\textsubscript{4}S\textsubscript{8} forms in a non-centrosymmetric cubic $F\bar{4}3m ~(T_d^2)$ lacunar spinel structure, comprising tetrahedral V\textsubscript{4} clusters with local spin 1/2. 
It undergoes a structural Jahn-Teller transition upon cooling below \textit{T}\textsubscript{JT}~=~44~K to a polar rhombohedral $R3m ~(C_{3v}^{5})$ phase by stretching the vanadium-clusters along one of the four body diagonals \cite{Pocha:CoM-2000-12-10, Yadav:PhyB-2008-403-5-9, Ruff:SciAdv-2015-1-10, Wang:PRL-2015-115-20, Hlinka:PRevB-2016-94-6}.
Below T\textsubscript{C}~=~13~K, magnetic order sets in, with three neighboring magnetic phases: a cycloidal state, a SkL, and a ferromagnetic state.
Similar results on GaV\textsubscript{4}Se\textsubscript{8} \cite{Fujima:PRevB-2017-95-18, Bordacs:SR-2017-7-7584, Ruff:PRevB-2017-96-165119}, suggest that the coupling of ferroelectric and magnetic orders is universal for the whole lacunar spinel family \cite{Phuoc:PRL-2013-110-3, Abd-Elm:PRL-2004-93-12, Dorolti:JACS-2010-132-16, Kim:NatCom-2014-5-3988, Guiot:NatCom-2013-4-1722, Singh:PRL-2014-113-13, Pocha:CoM-2000-12-10, Ruff:SciAdv-2015-1-10,Ruff:PRevB-2017-96-165119}.

In GaV\textsubscript{4}S\textsubscript{8}, at the Jahn-Teller transition, a multi-domain state with submicron-thick sheets of ferroelectric domains separated by \{100\}-type domain walls is created \cite{Kezsmarki:NatM-2015-14, Butykai:SR-2017-7-44663}, forming twinned areas with alternating orientation of the rhombohedral axis resulting in lamellar patterns \cite{Butykai:SR-2017-7-44663}.
The unconventional Jahn-Teller nature of the ferroelectricity \cite{Ruff:SciAdv-2015-1-10, Fujima:PRevB-2017-95-18, Singh:PRL-2014-113-13} implies entanglement between the ferroelectric and magnetic domains, imposes geometrical constraints on the SkL \cite{Butykai:SR-2017-7-44663}, and potentially opens possibilities to use these domain walls as active device elements \cite{Meier,Catalan12,Ruff17}.

In this context, it is interesting to explore the structural transition and domain-wall patterns of the lacunar spinel GaMo\textsubscript{4}S\textsubscript{8} \cite{Barz:MRB-1973-8-8}.
While in this material the Jahn-Teller-active Mo\textsubscript{4} units have the same spin of 1/2 as the V\textsubscript{4} clusters in the vanadium based lacunar spinels, the highest molecular orbital of the Mo\textsubscript{4} tetrahedra is occupied by one hole instead of one unpaired electron of the V\textsubscript{4} tetrahedra, i.e. the electronic configuration of GaMo\textsubscript{4}S\textsubscript{8} is different.
As a result, below the cubic to rhombohedral transition, taking place at \textit{T}\textsubscript{JT}~=~47~K in GaMo\textsubscript{4}S\textsubscript{8}, the spontaneous strain develops with opposite sign in this compound: its lattice is contracted along the polar rhombohedral axis\cite{Pocha}, whereas it is elongated in GaV\textsubscript{4}S\textsubscript{8}\cite{Pocha} and GaV\textsubscript{4}Se\textsubscript{8}\cite{Bichler2010}.
The structural transition in GaMo\textsubscript{4}S\textsubscript{8} is followed by  a magnetic transition at \textit{T}\textsubscript{c}~$= 18.5$~K, as indicated by specific heat, magnetization and magnetocaloric measurements \cite{Shamrai1982}.
Metamagnetic phase transitions were detected in both GaMo\textsubscript{4}S\textsubscript{8} and GaMo\textsubscript{4}Se\textsubscript{8} powder crystalline samples by magnetization and differential susceptibility measurements \cite{Rastogi:bssp-1987-142-2, Rastogi1983}.
Based on symmetry arguments and the analogy of this compound to GaV\textsubscript{4}S\textsubscript{8} and GaV\textsubscript{4}Se\textsubscript{8}, we may assume that the H=0 ground state of GaMo\textsubscript{4}S\textsubscript{8} is a cycloidally modulated spin ordering.

In our study, we have combined several complementary scanning probe microscopy techniques, including non-contact atomic force microscopy (nc-AFM), scanning dissipation microscopy (SDM), frequency-modulated Kelvin-probe force microscopy (KPFM), and out-of-plane piezoresponse force microscopy (PFM), in order to investigate the essential aspects of the multi-domain state formed below the Jahn-Teller transition in GaMo\textsubscript{4}S\textsubscript{8}. 
We have found that the lamellar patterns of $\{100\}$-type ferroelectric domain walls emerge with a typical domain thickness of $\sim$~25~nm, two orders of magnitude smaller than in GaV\textsubscript{4}S\textsubscript{8}.
Besides the lamellar patterns of uncharged primary domain walls, we also observed two types of secondary domain walls some of which are electrically charged, with head-to-head and tail-to-tail orientation of the polarization.
Moreover, we have identified a topological line defect, at which the electrically neutral secondary domain wall, connecting a quadruplet of primary domain states, branches into two charged secondary domain walls, associated with triplets of primary domain states.
\section{Experimental}
\subsection{Macroscopic properties}
\begin{figure}[hb]
\includegraphics[width=\columnwidth]{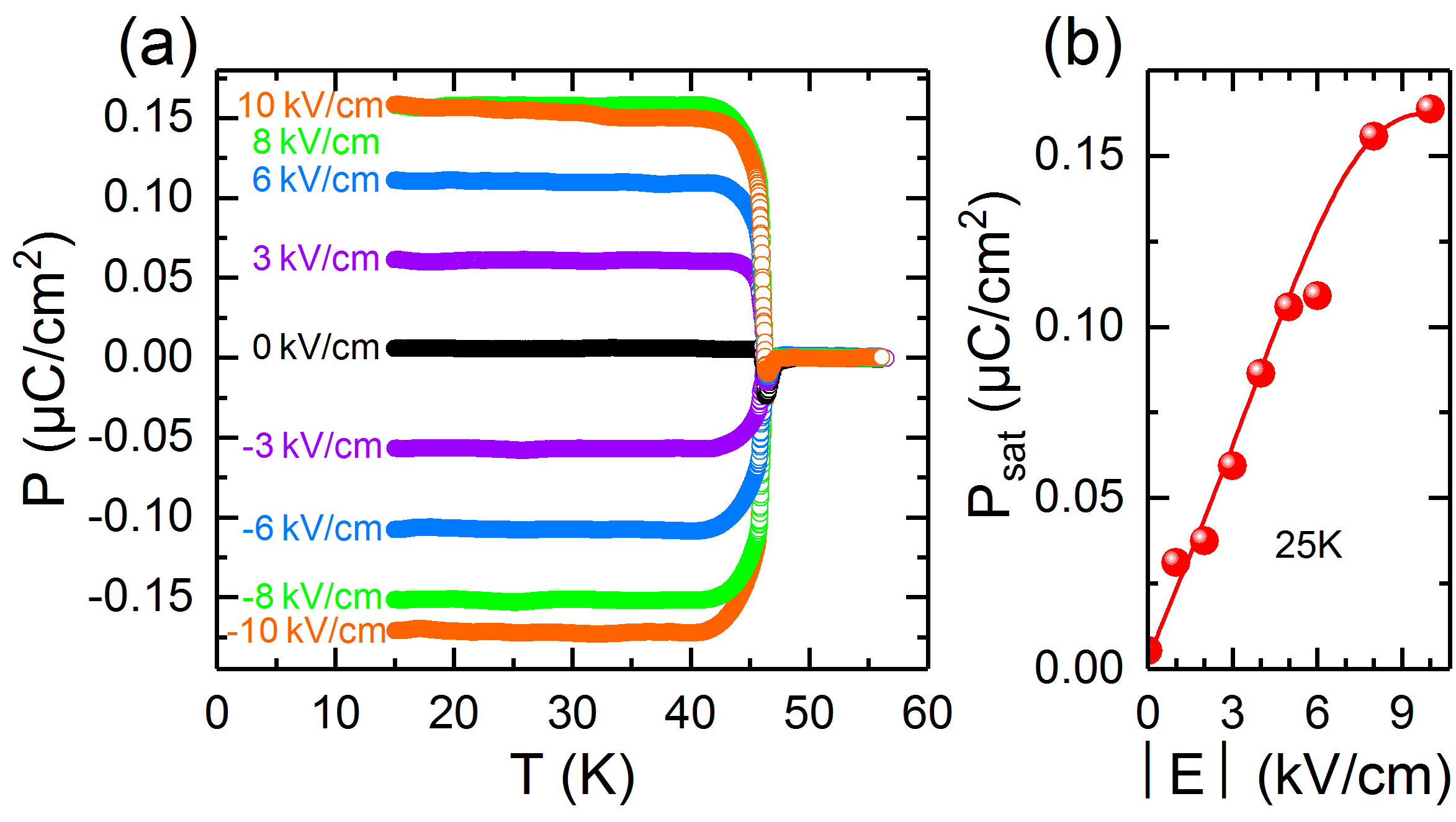}
 \caption{\label{fig1} 
 (a) Temperature dependence of the polarization of GaMo\textsubscript{4}S\textsubscript{8} measured during heating along the crystallographic $[110]$ direction for different poling fields applied along the same axis.
 (b) The magnitude of the polarization at 25~K, $P_{\textrm{sat}}=[P(E)-P(-E)]/2$, \textit{vs} the absolute value of the poling field. 
 The line is a guide for the eye. Relative experimental uncertainties are about the twice of the size of the data symbols, but the saturation of the spontaneous polarization at high poling fields is clearly apparent also from the raw data shown in panel (a).
 }
\end{figure}

The macroscopic polar properties of GaMo\textsubscript{4}S\textsubscript{8} were studied by pyroelectric current measurements for different poling electric fields along the $[110]$ axis.
Fig.~\ref{fig1}(a) shows the temperature dependence of the  polarization ${\bf P}$ projected along the same axis, as obtained by integrating the pyrocurrent versus temperature during heating in the absence of external electric field (see experimental details in the methods section).
Upon the Jahn-Teller transition, taking place at \textit{T}\textsubscript{JT}~=~47~K, spontaneous polarization drops down rather abruptly, implying a first order transition, similar to that of GaV\textsubscript{4}S\textsubscript{8}\cite{Ruff:SciAdv-2015-1-10}, GaV\textsubscript{4}Se\textsubscript{8}\cite{Ruff:PRevB-2017-96-165119}, and GeV\textsubscript{4}S\textsubscript{8}\cite{Singh:PRL-2014-113-13}.
By moderate electric poling fields, the direction and the magnitude of polarization can be controlled, proving polarization switching and, thus, the ferroelectric nature of the polar phase in GaMo\textsubscript{4}S\textsubscript{8}. 
As shown in Fig.~\ref{fig1}(b), the saturated polarization $P_{\rm sat}$ first increases linearly with increasing strength of the poling field and than becomes more flat around $\approx$~9~kV/cm, approaching a value of approximately $0.15~\mu\textrm{C}/\textrm{cm}^2$. 
This likely indicates that the multi-domain state involving primary domains with polarization along the four $\langle 111 \rangle$-type axes [see Fig.~\ref{fig3}(a)] is transformed to the polar mono-domain state favoured by the poling field.
However, due to the coplanar contact geometry and the fact that the voltage is applied along the [110] direction, oblique to the $[111]$ ferroelectric axis, we estimate that the actual spontaneous polarization is about $P\approx 0.2-0.4~\mu\textrm{C}/\textrm{cm}^2$
and the relative uncertainty of the data in panel (b) is about the twice of  the size of the data symbols.
This value is comparable to the magnitude of the polarization observed in GaV\textsubscript{4}S\textsubscript{8} and GaV\textsubscript{4}Se\textsubscript{8}\cite{Ruff:SciAdv-2015-1-10, Ruff:PRevB-2017-96-165119}.
Let us stress that the switching was so far achieved only in the field-cooled protocol described above.

\subsection{Microscopic properties}
\begin{figure*}[ht]
  \includegraphics[width=\textwidth]{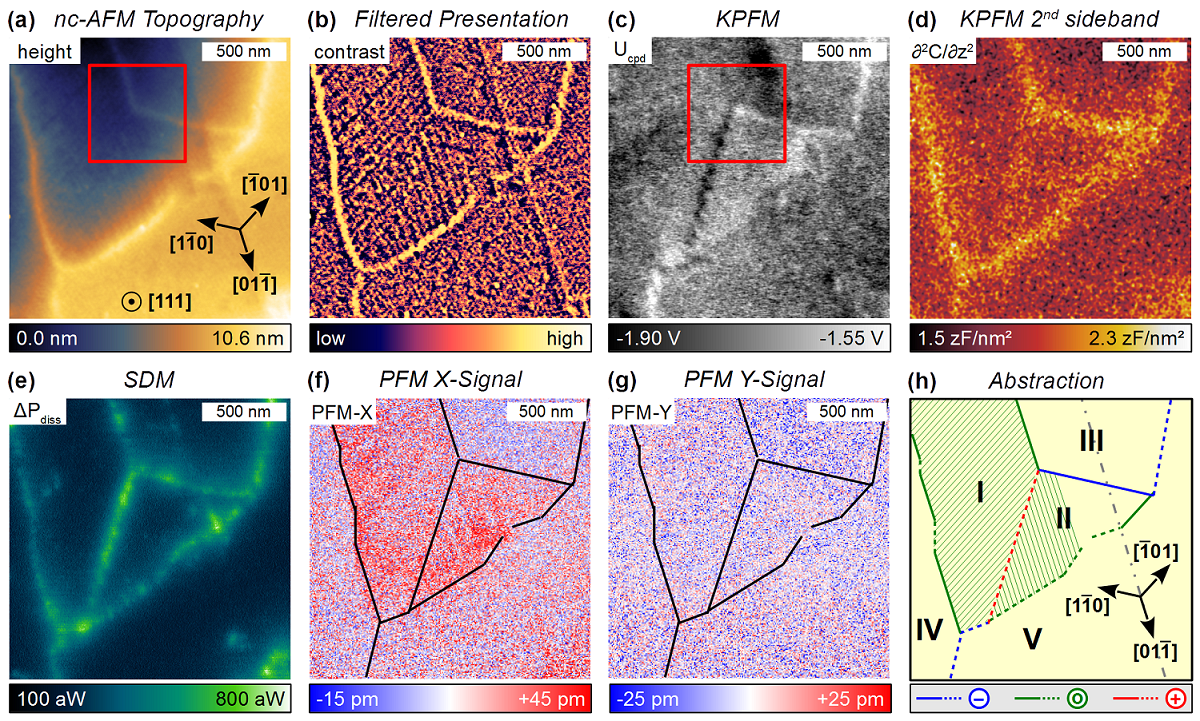}
  \caption{\label{fig2} 
  Typical real-space images of the surface structures on GaMo\textsubscript{4}S\textsubscript{8} obtained by SPM for the $(111)$ surface.
  (a-e) Non-contact topography, KPFM, KPFM second-sideband and SDM images measured at \textit{T}~=~7.8~K.
  Filtered presentation of the topography in panel b emphasizes the surface curvature, the red square in panels a and c marks the area enlarged in Fig.~\ref{fig4}(a,b).  
  (f,g) Out-of-plane PFM images recorded over the same region at \textit{T}~=~11.1~K.
  The black lines represent the features visible in panel e.
  (h) Schematic summary of the observed features. The lines correspond to topographic features visible in panel b, while their color encodes the sign of the charging according to panel c.
  }
\end{figure*}

In Fig.~\ref{fig2} we present a series of typical images obtained on the $(111)$ surface of an as-grown GaMo\textsubscript{4}S\textsubscript{8} single crystal at temperatures well below \textit{T}\textsubscript{JT}. 
Panels (a-e) summarize the simultaneously measured topography, KPFM, KPFM second sideband, and SDM at \textit{T}~=~7.8~K.
Arrows in Fig.~\ref{fig2}(a) indicate the orientation of the three in-plane $\langle110\rangle$-axes, as determined  by x-ray diffraction.
In Fig.~\ref{fig2}(f,g), we show PFM images obtained at \textit{T}~=~11.1~K over the same sample area. 

The topography image depicted in Fig.~\ref{fig2}(a) reveals several featureless flat areas with lateral dimensions ranging from 500~nm to several {\textmu}m. They are mutually slightly inclined, forming folds in the surface.
However, two of the areas are puckered resulting in a clearly visible stripe pattern.
For better visibility of these lamellar patterns, a filtered presentation of the topography emphasizing the surface curvature is presented in Fig.~\ref{fig2}(b).
The lamellas are typically 30~nm to 60~nm wide, show an average height modulation of $\approx 2$~{\AA}, and run along one of the in-plane $\langle110\rangle$ axes. 
Both, the surface puckering as well as the surface folds disappeared upon increasing the temperature above \textit{T}\textsubscript{JT}, and hence, they can clearly be attributed to the formation of ferroelectric domains.
Measuring the width of more than 900 such lamellar domains in this and many independent images and considering the domain wall inclination with respect to the sample surface (a factor of $\sqrt{3/2}$, see below), we find the most probable lamellar domain width in this sample to be $d_{DW} \approx 25$~nm, which is roughly three times larger than the expected magnetic modulation length\cite{NakamuraPrivate}. 

The resulting schematic image in Fig.~\ref{fig2}(h) shows the separate plane areas, numbered with capital roman numbers from I to V for ease of discussion.
Herein, thicker lines illustrate the observed folds in topography with continuous and dashed lines indicating folds parallel and oblique to one of the in-plane $\langle110\rangle$ axes, respectively. 
The two specific areas with stripes are denoted I and II.
The gray dash-dotted line refers to a surface groove visible in topography even above \textit{T}\textsubscript{JT}, and, therefore, it is not related to a ferroelectric domain wall.

In order to compensate electrostatic interactions, a dc-voltage of $-106$~V was applied to the tip, while the KPFM loop controlled the dc-bias applied to the sample.
The corresponding contact potential difference is depicted in Fig.~\ref{fig2}(c).
The line separating areas I and II depicts a negative (dark) contrast, which corresponds to a positive charging relative to the adjoined areas.
Accordingly, a positive (bright) contrast corresponds to a negative charging relative to the surroundings. 
These line features are depicted in the schematic Fig.~\ref{fig2}(h) as red (positive) and blue (negative) lines.
In contrast, neither the fine stripe patterns in areas I and II, nor the borderline separating areas I and III are visible in the contact potential difference.

In Fig.~\ref{fig2}(d), we show a map of the amplitude of the so-called second KPFM sideband.
Running the KPFM control loop, a modulation voltage is applied which leads to two sidebands around the resonance frequency in the spectrum of the cantilever motion.
The amplitude of the second sideband carries information about the second derivative (curvature) of the dependence of the effective tip-sample system  capacitance on the tip-sample distance \cite{Zerweck:PRB-2005-71-12,Garrett16}.
All folds visible in topography feature a higher second-sideband amplitude, i.e. for the tip approaching to the sample the raise of the tip-sample capacitance is larger at the positions of the folds.
In the dissipation signal, shown in Fig.~\ref{fig2}(e), also a higher dissipation is found at the positions of the folds.
Finally, in Fig.~\ref{fig2}(f,g) we present the out-of-plane PFM-signal obtained over the same area with the black lines indicating the positions of the folds.
In areas I and II stripes are visible in the out-of-plane PFM X-signal [Fig.~\ref{fig2}(f)], while all other areas appear uniform.
The magnitude of the observed inverse piezoelectric effect is about 1-2~pm/V.

\section{Symmetry analysis}
\begin{figure*}[ht]
  \includegraphics[width=\textwidth]{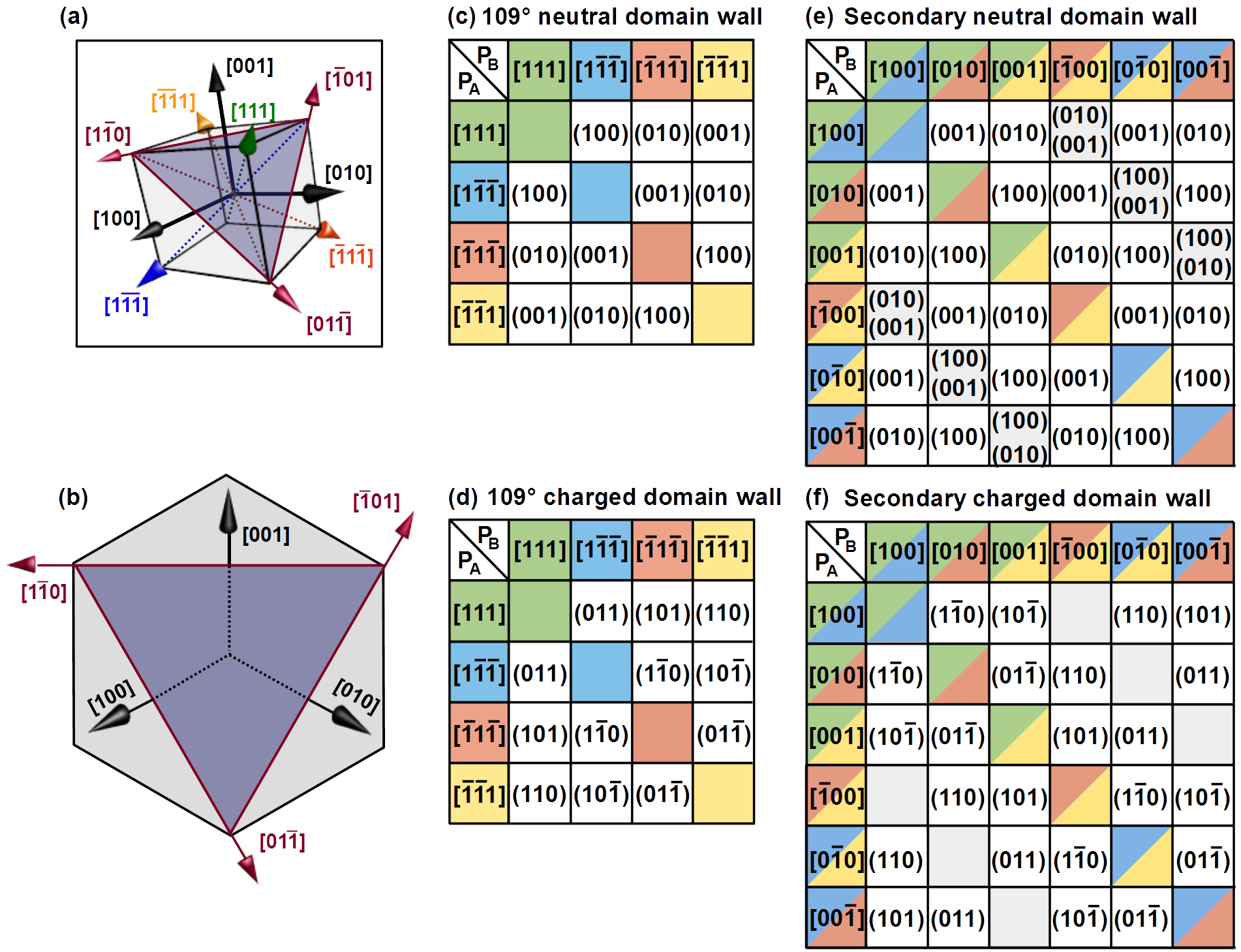}
  \caption{\label{fig3} 
  Ferroelectric domain states and mechanically compatible domain walls associated with the $\bar{4}3m>3m$ macroscopic symmetry reduction. 
  (a) The four $\langle 111 \rangle$-type directions of the spontaneous polarization within the parent cubic reference frame.
  Selected face diagonals parallel to $[01\bar{1}]$, $[\bar{1}01]$ and $[1\bar{1}0]$ directions define the $(111)$ plane, parallel to the sample surface.
  (b) View along the $[\bar{1}\bar{1}\bar{1}]$ direction. 
  (c,d) Crystallographic orientation of the  primary neutral and charged domain walls, respectively.
  (e,f) Crystallographic orientation of the secondary neutral and charged domain walls, respectively.
  Distinct domain states are labeled by the average polarization direction in the top row and the left column, the inner fields indicate the permissible orientations of the mechanically compatible interfaces.
  }
\end{figure*}

Before proceeding with the explicit assignment of the observed patterns, let us briefly summarize the domain properties implied by the $\bar{4}3m>3m$ macroscopic symmetry reduction.
It is known \cite{species} that this symmetry breaking species (No.\,172) yields four distinct macroscopic domain states, uniquely distinguishable by the orientation of their spontaneous polarization which can be labeled as $[111]$, $[1\bar{1}\bar{1}]$, $[\bar{1}1\bar{1}]$, and $[\bar{1}\bar{1}1]$ in relation to the parent cubic reference [Fig.~\ref{fig3}(a)], here called primary domains.
The corresponding domain pairs can be separated by 109-degree ferroelectric domain walls.
Since the spontaneous off-diagonal strain tensor elements scale with the spontaneous polarization, $(\epsilon_{yz},\epsilon_{zx}, \epsilon_{xy}) \propto (P_{x},P_{y}, P_{z})$ and the diagonal elements are all equal,
mechanical compatibility implies, that for each domain pair $\{\bf{P}_{\rm A},\bf{P}_{\rm B} \}$ there is one mechanically compatible wall orientation with its wall normal along $ \bf{P}_{\rm A} +\bf{P}_{\rm B} $, and another one with its domain wall normal along $ \bf{P}_{\rm A} -\bf{P}_{\rm B} $.
The former one is an electrically neutral wall, while the latter is a charged wall \cite{Fousek}.
Crystallographic orientations of these primary domain walls are summarized in Fig.~\ref{fig3}(c,d).

Twinned areas typically involve only two primary domain states, separated by a sequence of parallel domain walls, which can be considered as a macroscopic super-domain, here called a secondary domain.
Assuming equal volume of the two primary states and a negligible macroscopic impact of the primary domain walls involved, we arrive to six secondary domain variants, uniquely distinguishable by their average polarization ($[100]$, $[010]$, $[001]$, $[\bar{1}00]$, $[0\bar{1}0]$, and $[00\bar{1}]$).  
Inspection of the off-diagonal elements of the average strain tensor in the secondary domains allows to enumerate mechanically compatible secondary domain walls, which can be treated as domains of the $\bar{4}3m>m_{\backslash}m2_{+}$ symmetry reduction (see species No.~176 in Ref. \cite{species}).
Both 180 degree and 90 degree ferroelectric secondary walls are possible.
In case of the 90-degree secondary domain pair, such as $[100]$ and $[010]$ (or $[100]$ and $[0\bar{1}0]$), one mechanically compatible wall is electrically neutral and the other one is charged, while both mechanically compatible 180-degree domain walls are neutral ones.
The normal of neutral and charged 90-degree walls is given by the cross product and the difference of the adjacent secondary domain polarizations, respectively.
A summary of the orientations of all mechanically compatible secondary walls is given in Fig.~\ref{fig3}(e,f).
Moreover, it is easy to demonstrate that secondary domains can only be compatible with those primary domain states that are contained in this secondary domain, and that the corresponding mechanically compatible interfaces are those already listed in Fig.~\ref{fig3}(c,d).

The surface traces of planar domain walls are directed along surface vectors ${\bf s}_{\rm S} = {\bf n} \times {\bf s}$, where ${\bf n}$ and ${\bf s}$ are the sample surface normal unit vector and the domain wall normal unit vector, respectively.
Similarly, the projection of the polarization on the surface normal reads ${\bf P}_{\rm \perp} = {\bf n} ( {\bf n} \cdot {\bf P} )$.
In case of the $(111)$ surface, ${\bf P}_{\rm \perp}$ is unique for the $[111]$ but equal for any of the obliquely incident $[1\bar{1}\bar{1}]$, $[\bar{1}1\bar{1}]$, and $[\bar{1}\bar{1}1]$ primary domains and the longitudinal out-of-plane piezoelectric response cannot distinguish among these three states \cite{Butykai:SR-2017-7-44663}.

In addition, perfect crystallographic matching of the adjacent ferroelastic domains requires a rotation of the lattice planes, and this puckers of the originally flat outer surface of the twinned crystal.
On the $(111)$ surface, neutral primary domain walls induce such a puckering only if the $[111]$ state is involved. Pursuant to the argumentation used for GaV\textsubscript{4}S\textsubscript{8} in Ref. \cite{Butykai:SR-2017-7-44663},  and using the reported spontaneous strain magnitude of GaV\textsubscript{4}S\textsubscript{8} \cite{Powell}, $|\epsilon_{xy}| =0.007$, the surface puckering angle should be about 0.8~degree.

Finally, it is worth mentioning, that a priori we do not know whether the investigated crystal surface is oriented in a way, such that domains with their polarization normal to the surface are pointing outward or inward.
In the following, we shall systematically assume it is the outward direction, as suggested in Fig.~\ref{fig3}.
In the opposite case, all the polarization directions would change to the opposite ones as well as the nominal domain-wall charge densities $Q$ due to the divergence of the spontaneous polarization, which calculate to: $Q= {\bf P}_{\rm A} \cdot {\bf s}_{\rm A}+{\bf P}_{\rm B} \cdot  {\bf s}_{\rm B} $.
Here, ${\bf s}_{\rm A}$, $ {\bf s}_{\rm B} $ are the domain-wall normal unit vectors oriented from the inside of the respective domain towards the domain-wall location.
Yet, the orientation of all mechanically compatible walls indicated in Fig.~\ref{fig3} as well as their intersections with the surface would remain intact by the inversion operation.
\section{Discussion}
\begin{figure}[ht]
  \includegraphics[width=\columnwidth]{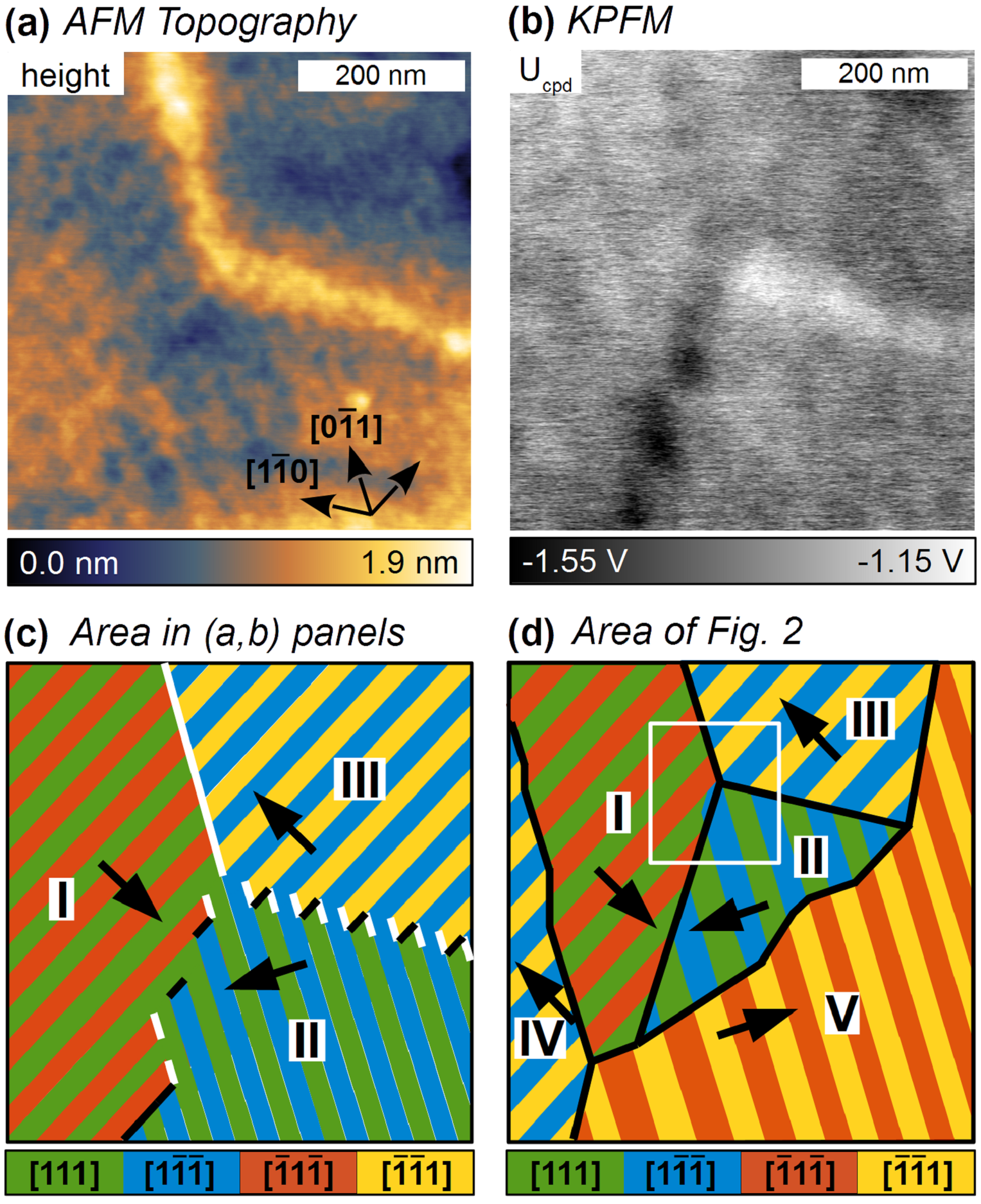}
  \caption{\label{fig4}
  Analysis of the domain distribution in GaMo\textsubscript{4}S\textsubscript{8}.
  (a,b) Topography and KPFM images within zoomed area marked in Fig.\,\ref{fig2}(a,c). 
  (c) Attempt to assign secondary domains of panels (a,b) yielding locally incompatible domain wall configurations at macroscopically mechanically compatible secondary domain walls.
  The colors depict primary domain states while the line colors represent the sign the domain wall polarization charge density.
  Arrows indicate the in-plane projected secondary domain polarization.
  Panel (d) presents the final identification of the observed secondary domains in Fig.~\ref{fig2} and panels a,b (white square).
  }
\end{figure}

The considerations presented above allow to understand all major features of the observed domain structures.
Areas I and II show obvious surface puckering stripes, revealing nicely the expected primary twinning [see Fig.~\ref{fig2}(b,h)].
The surface puckering implies that one of the primary domain states involved is the $[111]$ state, and the stripe orientation evident from the topography images equals the crystallographic orientation of the intersections of the involved primary domain walls with the $(111)$ surface.
Furthermore, we can conservatively assume that the primary twinning is due to the electrically neutral walls, which is at least partly corroborated by the fact that these stripes are not seen in the KPFM image, neither here in GaMo\textsubscript{4}S\textsubscript{8} nor in the GaV\textsubscript{4}S\textsubscript{8} homologue crystal, where these primary walls were observed with a larger spacing. 
Therefore, area I involves $[111]$ and $[\bar{1}1\bar{1}]$ states, and area II involves $[111]$ and $[1\bar{1}\bar{1}]$ states, corresponding to the secondary domains labeled as $[010]$ and $[100]$, respectively. 
The assignment is marked in Fig.~\ref{fig4}(c,d) by the arrows indicating the in-plane projection of the secondary domain polarization ${\bf P}_{||} = {\bf P} - {\bf n} ( {\bf n} \cdot {\bf P} )$ as well as by the colored stripes indicating the primary domains and the orientation of the surface traces of the primary domain walls.

The  mechanically compatible walls between the secondary domains $[010]$ and $[100]$ can be either $(001)$ or $(1\bar{1}0)$ planes [see Fig.~\ref{fig3}(e,f)].
Apparent in Fig.~\ref{fig2}, the borderline between areas I and II corresponds to the latter case implying a charged head-to-head domain wall, which is nicely confirmed by the marked contrast in the KPFM image [see Fig.~\ref{fig4}(b) for details].

The  area III is adjacent to both areas I and II [see Fig.~\ref{fig4}(c,d)].
The only primary domain state, mechanically compatible with both areas simultaneously, is the $[111]$ domain state, but the compatible interfaces with this state would require its borderlines being parallel to the stripes in the neighboring secondary domains, which was not observed.
Therefore, area III has to be a secondary domain. 
Out of the three secondary domains without surface puckering, only the $[0\bar{1}0]$ secondary domain fulfills the simultaneous mechanical compatibility.
Consequently, the III~--~II borderline corresponds to the $(110)$ oriented 90-degree charged wall with tail-to-tail configuration, while the III~--~I borderline corresponds to the $(100)$ oriented 180-degree neutral wall.
These assignments are well supported by the almost vanishing and by the finite negative KPFM signal, respectively [see Figs.~\ref{fig2}(c) and \ref{fig4}(b)], as well as by the out-of-plane PFM image of Fig.~\ref{fig2}(f), which shows distinct signal in the puckered and not-puckered surface areas.
 
\begin{figure}[ht]
  \includegraphics[width=\columnwidth]{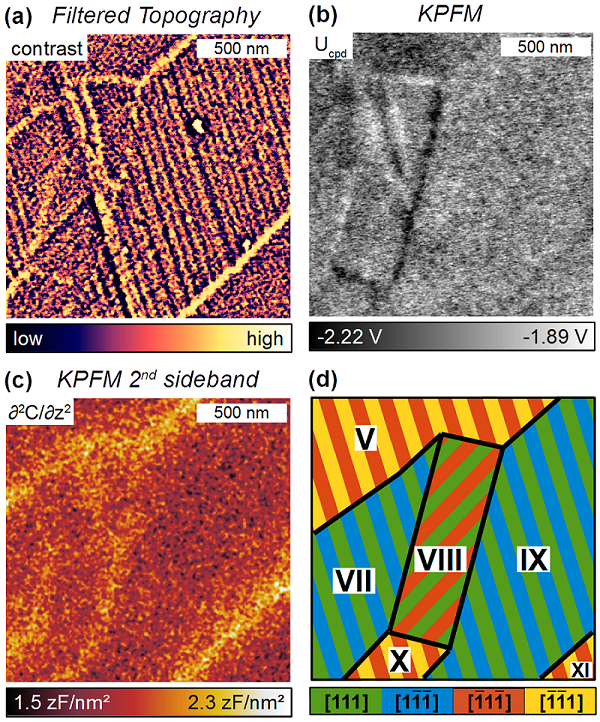}
  \caption{\label{fig5}
  Domain distribution in another area of the GaMo$_4$S$_8$ crystal: (a, b, c) Non-contact topography, KPFM, and KPFM second-sideband images at $T = 7.8$\,K. Panel (d) presents the final identification of the observed domains. Note that the area V is continuation of the same area V in Figs.~\ref{fig2} and ~\ref{fig4}d.
  }
\end{figure}

Similarly, it is possible to uniquely assign the remaining areas IV and V in the square image of Fig.~\ref{fig2} [see Fig.~\ref{fig4}(d)].
In particular, all straight boundaries in the resulting domain configuration, sketched in Fig.~\ref{fig4}(d), correspond to orientations of mechanically compatible secondary domain walls, except for a few very short minor segments.
The above described domain architecture is rather typical in the investigated sample, another example of a similar domain arrangement is shown in Fig.~\ref{fig5}.

The thickness of the primary domain walls could not be determined from the present experiment, but since the phase transition dynamics is of strongly order-disorder type \cite{Hlinka:PRevB-2016-94-6}, it can be expected that they are actually atomically narrow. 
On the other hand, the apparent thickness of the secondary domain walls (such as in Fig.~\ref{fig4}) is about 30 nm, i.e. comparable with the thickness of the primary domains themselves.
This apparent thickness of secondary domain walls is visibly larger than the typical thickness of primary ferroelastic domain walls in ordinary ferroelastic materials \cite{Chrosch99}.
Notably, the compatibility of adjacent secondary domains does not imply whether the secondary wall between them is perfectly planar also at the nanoscale or whether it is formed as a zig-zag wall, such as the wall sketched in Fig.~\ref{fig4}(c). 
Some hints about the possible nanoscale architecture can be inferred from the compatibility of the primary domains.
For example, the matching between the $[111]$ and $[\bar{1}\bar{1}1]$ primary domains and the $[1\bar{1}\bar{1}]$ and $[\bar{1}1\bar{1}]$ primary domains at the $(100)$ oriented I--III secondary wall, sketched in Fig.~\ref{fig4}(c), would imply an electrically and mechanically incompatible connection in the whole area of this domain wall.
Obviously, it is much more likely that this secondary wall is rather formed by the $(100)$ oriented, electrically neutral and mechanically compatible primary domain wall segments between $[111]$ and $[\bar{1}1\bar{1}]$ or $[1\bar{1}\bar{1}]$ and $[\bar{1}\bar{1}1]$ primary domains, as suggested in Fig.~\ref{fig4}(d).
Still, it should be mentioned that there are four different primary domains joint at certain lines within such a wall, and even though neighboring domains form compatible pairs, the whole domain quadruplet is not fully compatible \cite{Mokry:JoAP-2005-97-11} and probably additional strain arises to suppress the surface puckering in the vicinity of such a secondary wall.

Similar reasoning can be made for the charged 90 degree wall at the II--III interface.
In principle, it can be formed by mechanically compatible primary walls parallel to the secondary domain wall orientation alternated by simply connected $[1\bar{1}\bar{1}]$ primary domain regions, but this would result in a very uneven charge distribution.
More likely is a zig-zag wall of equally charged mechanically compatible primary wall segments.
In the real material, various imperfections of the nanoscale arrangement can be expected and might be even responsible for the fluctuations of the KPFM signal seen in Fig.~\ref{fig4}(b).

The relatively low piezoelectric response of the sample and natural limitations imposed by the liquid-helium cryogenic environment prevent us from application of complementary scanning techniques like PFM imaging near contact resonance \cite{Rodriguez2007} and lateral or vector PFM \cite{Eng1998, Kalinin2006, Nath2010} or angle-resolved PFM \cite{Park2010}.

\section{Conclusion}
From a fundamental standpoint, perhaps the most interesting result is the observation of junctions connecting triplets of secondary domain walls, such as those between areas I, II, and III in Fig.~\ref{fig4}(d).
These mechanically compatible secondary walls can meet at a single line, parallel to the $[100]$ direction, where at the same time, up to six primary domains can meet in the core of this linear defect.
Motion of this linear defect realizes zipping or unzipping of the electrically neutral wall into two oppositely charged walls.
For example, by applying a suitable external force favoring the domain state of area II at the expense of areas I and III, the topological defect acts as a charge-separating nanodevice.

To sum up, we observed a switchable ferroelectric polarization of $\gtrsim$~0.20~{\textmu }C/cm\textsuperscript{2} in the rhombohedral phase of GaMo\textsubscript{4}S\textsubscript{8}, and imaged the corresponding ferroelectric-ferroelastic domains using a set of scanning probe microscopy techniques.
Using PFM, the magnitude of the effective piezoelectric coefficient within the $[111]$ primary ferroelectric domain was determined to be $\approx 2$~pm/V.
Combining scanning probe microscopy and theoretical approaches, we found that the domain structure  is composed of secondary domains, formed by a lamellar pattern of primary ferroelectric domains separated by $\{100\}$-type domain walls.
Some of the secondary domain walls are neutral, some of them are charged.
Furthermore, the topological line defects observed at the junctions of three secondary domain walls can be exploited to reconfigure the domain wall structure on the nanoscale, e.g. by transforming neutral walls to charged ones.

The typical width of the observed lamellar domains ranges between 10\,nm and 100\,nm,  with the maximum of the domain width distribution at $\approx 25$~nm.
Since the q-vector of modulated magnetic textures, like cycloids and skyrmion lattices, are restricted to the plane perpendicular to the polar axis in the present crystal symmetry \cite{Kezsmarki:NatM-2015-14, Butykai:SR-2017-7-44663, Bogdanov:ZETF-1989-95-1}, the small width of the lamellar domains severely limits the lateral extension of such textures.
Such geometrical confinement, reported to be an efficient tool to stabilize and control skyrmions in metallic skyrmion hosts\cite{Jin2017, Zhao2016}, is a subject of future studies on GaMo\textsubscript{4}S\textsubscript{8}.\\
\ack
P.M., E.N., and L.M.E. gratefully acknowledge financial support by the German Science Foundation (DFG) through the Collaborative Research Center ``Correlated Magnetism: From Frustration to Topology'' (SFB 1143) and grant numbers MI~2004/3-1 and EN~434/38-1 as well as by the German Academic Exchange Service (DAAD) through grant numbers 57333728 and 57336108.
P.O. and J.H. gratefully acknowledge financial support  by  the Czech Science Foundation (project no. 17-11494J).
K.G., P.L. and I.K. gratefully acknowledge financial support by the DFG via the Transregional Research Collaboration TRR80: From Electronic Correlations to Functionality (Augsburg/Munich/Stuttgart).
\section*{Appendix}
\appendix 
Single crystals of GaMo$_4$S$_8$ were grown by the flux method in a sealed molybdenum tube \cite{Querre}.
For the polarization measurement, the crystal platelet was placed into a \textsuperscript{4}He-bath cryostat (Oxford).
Two coplanar 0.44~mm long contacts of silver paint were applied on the same as-grown surface of the crystal.
The resulting 0.1~mm gap was perpendicular to the crystallographic $[110]$ direction.
The sample was 0.25~mm thick, but for calculating the current density we assumed a penetration depth of the electric field equal to the contact distance, i.e., 0.1~mm. 
The electric field was applied at 65~K and consequently the sample was cooled with about 5~K/min down to 4~K.
Then, the the poling field was switched off and the sample was connected to a Keithley 6517 electrometer to record the pyrocurrent upon reheating the sample up to 60~K with 5~K/min.
Integrating the pyrocurrent density, the change of polarization during the heating process was deduced. 

The crystal used for our scanning probe study, was first cleaned in an ultrasonic acetone bath and thereafter the surface was cleaned mechanically with a CO\textsubscript{2}-snow jet.
For electric contact, the sample was contacted with conductive silver paint.
We used an Omicron low-temperature ultra-high vacuum atomic force microscope equipped with RHK Technology Inc. R9-control electronics using Nanosensor SSS-QMFMR probes (spring constant $k \approx 2.9$~N/m, resonance frequency $f_{0} \approx 74.5$~kHz).
The Q-factor was enhanced by means of the interaction with the microscopes interferometer cavity to effectively $Q_{\rm eff} \approx 2\cdot 10^5$.
Non-contact measurements were done at a frequency shift of $\Delta f = -60$~Hz, the oscillation amplitude was controlled to 10~nm, and local differences of the contact potential were compensated by running a Kelvin-probe force controller.
For KPFM we used a modulation voltage $U_{mod} = 2$~V with a frequency of $4.1$~kHz.
Later, the same tip was used for out-of-plane PFM with a PFM excitation voltage of $U_{\rm mod} = 10$~V at a frequency of $6.1$~kHz applied to the tip, while the sample was grounded.
Filtering of the topography for emphasized curvature contrast, Fig.\,\ref{fig2}(b), was achieved with the SEM image presentation tool of the Gwyddion software \cite{GwyMan}.
\section*{References}
\bibliography{bibliography}
\bibliographystyle{unsrt}
\end{document}